\documentclass[aps,twocolumn,nofootinbib,showpacs,floatfix, superscriptaddress]{revtex4}
\usepackage{graphics}
\usepackage{amssymb,amsmath}
\usepackage{amsmath}
\usepackage{psfrag}
\usepackage{graphicx}
\newcommand{\bx}{{\bf x}}
\newcommand{\bR}{{\bf R}}
\newcommand{\bk}{{\bf k}}
\newcommand{\bu}{{\bf u}}
\newcommand{\bH}{{\bf H}}
\newcommand{\s}{\sigma}

\newcommand{\be}{\begin{equation}} 
\newcommand{\ee}{\end{equation}} 
\newcommand{\bea}{\begin{eqnarray}} 
\newcommand{\eea}{\end{eqnarray}}  
\newcommand{\bean}{\begin{eqnarray*}} 
\newcommand{\eean}{\end{eqnarray*}}

\begin{document}

\title{Tilings of space and superhomogeneous point processes}

\author{A. Gabrielli}
\affiliation{SMC-INFM, Dipartimento di Fisica, 
Universit\`a ``La Sapienza'', P.le
A. Moro 2, I-00185, Rome, Italy}
\affiliation{ISC-CNR, Via dei Taurini 19, I-00185
Rome, Italy}

\author{M. Joyce} 
\affiliation{Laboratoire de Physique
Nucl\'eaire et de Hautes Energies, UMR-7585 \\
Universit\'e Pierre et Marie Curie --- Paris 6, 
75252 Paris Cedex 05, France}

\author{S. Torquato} 
\affiliation{Department of Chemistry, Princeton University, Princeton,
NJ 08544, USA}

\affiliation{Program in Applied and Computational Mathematics, Princeton
University, Princeton NJ 08544}

\affiliation{Princeton Institute for the Science and Technology of Materials, Princeton University, 
Princeton, NJ 08544, USA}

\affiliation{Princeton Center for Theoretical Physics, Princeton University, Princeton,
NJ 08544, USA}

\begin {abstract}
\begin{center}
{\large\bf Abstract}
\end{center}
We consider the construction of point processes from tilings, with
equal volume tiles, of $d$-dimensional Euclidean space $\mathbb{R}^d$.
We show that one can generate, with simple algorithms ascribing one or
more points to each tile, point processes which are
``superhomogeneous'' (or ``hyperuniform''), i.e., for which the
structure factor $S({\bf k})$ vanishes when the wavenumber $k$ tends
to zero. The {\it exponent} of the leading small-$k$ behavior, $S(k
\rightarrow 0) \propto k^\gamma$, depends in a simple manner on the
nature of the correlation properties of the specific tiling and on the
conservation of the mass moments of the tiles.  Assigning one point to
the center of mass of each tile gives the exponent $\gamma=4$ for 
any tiling in which the shapes
and orientations of the tiles are short-range correlated. Smaller
exponents, in the range $4-d<\gamma<4$ (and thus always
superhomogeneous for $d\leq 4$), may be obtained in the case that the
latter quantities have long-range correlations. Assigning {\it more
than one point} to each tile in an appropriate way, we show that one
can obtain arbitrarily higher exponents in both cases.  We illustrate
our results with explicit constructions using known deterministic
tilings, as well as some simple stochastic tilings for which we can
calculate $S({\bf k})$ exactly. Our results
provide, we believe, the first explicit analytical construction of 
point processes with $\gamma > 4$. Applications to condensed 
matter physics, and also to 
cosmology, are briefly discussed.
\end{abstract}
\pacs{02.50.-r, 61.43.-j, 98.80.-k}
\maketitle

\section{Introduction}

``Superhomogeneous" \cite{glasslike} or ``hyperuniform" \cite{To03}
point patterns in $d$-dimensional Euclidean space $\mathbb{R}^d$ are
defined to be those in which infinite wavelength density fluctuations
vanish. In other words, the structure factor (or power
spectrum) $S({\bf k})$ of the number density field at wave vector $\bf
k$ has the following behavior:
\begin{equation}
   \lim_{\bk \rightarrow 0} S(\bk) = 0.
\label{Equation4}
\end{equation}
This defining characteristic of superhomogeneity (or hyperuniformity) 
is tantamount to saying that the usual mean-square particle-number 
fluctuations increases less rapidly than  $R^{d}$ for large $R$, 
where $R$ denotes the linear
size of an observation window in $\mathbb{R}^d$ \cite{glasslike,To03}.  
Indeed, the magnitude of such local density fluctuations have been 
suggested as  a possible ``order metric" to quantify the degree of order (disorder)
of an arbitrary  point pattern \cite{To03}.
Any superhomogeneous point pattern can be seen as a 
typical configuration of a particular type of  ``critical" point
in that the {\it direct correlation} function (defined
through the Ornstein-Zernike relation) is long-ranged while the
pair correlation function is short-ranged \cite{To03}. Such remarkable 
behavior is
diametric to that seen in usual thermal critical points in which the
inverse is true, i.e., the pair correlation is long-ranged
and the direct correlation function is short-ranged.

Although it is clear that any periodic point pattern is
superhomogeneous, it is less obvious that statistically translationally and
even rotationally invariant random point patterns in $\mathbb{R}^d$ can
have this property. We now know of a variety of intriguing
translationally and rotationally invariant random point patterns that
are superhomogeneous, including the ground state of liquid
$^4\mbox{He}$~\cite{Fe54,Fe56,Re67}, maximally random jammed
hard-sphere packings~\cite{Do05}, certain one-component
plasmas~\cite{OCP-review,lebo_and_all,modOCP}, the matter distribution in
the Universe~\cite{glasslike, lebo_and_all}, and certain aperiodic
tilings \cite{lebo_and_all,To03,Ha06}.  An interesting application of
superhomogeneous point patterns in cosmology is in the preparation
of initial conditions for gravitational $N$-body simulations
\cite{book,lebo_and_all,modOCP,Ha06}. Superhomogeneous distributions also
appear in cosmology in the context of the determination of bounds,
first derived by Zeldovich \cite{zeldovich-k4}, on the mass fluctuations at
large scales generated by causal mechanisms (i.e. with physics
respecting the causal constraints of cosmological models). Indeed, we
note that in this context a simplified form of the analysis we develop
here of the small-$k$ behavior of the structure factor is often used
(see e.g. Ref.~\cite{peebles}).

It is desirable to develop both theoretical and computational methods
to generate a wide class of superhomogeneous random point patterns.
Recently, a collective coordinate approach \cite{Uc04,Uc06} has been
employed to numerically generate translationally invariant
superhomogeneous point processes. This procedure enables one to
produce point patterns that completely suppress density fluctuations
of modes for a positive range of wavenumbers around the origin.  In
Ref.~\cite{gabrielli_etal_04} an algorithm for generating
discrete processes in one dimension with superhomogeneous mass 
fluctuations has been
given (see also Ref.~\cite{fratzl_etal1991}). 
An analytical methodology to relate superhomogeneous point processes and 
Voronoi tilings of space has recently been proposed and studied in 
Ref.~\cite{Ga04}.

In this paper, we study
the construction of superhomogeneous 
point patterns starting from generic tilings of Euclidean
space $\mathbb{R}^d$ with equal volume tiles. We show how to
explicitly generate such point processes in which the structure factor
for small wavenumbers has the power-law form $S(k) \sim k^\gamma$ for positive
$\gamma$, where $k\equiv |{\bf k}|$ is the wavenumber.  The
constructions illustrate the very specific properties of these
superhomogeneous point patterns in which the exponent $\gamma$,
characterizing the long wavelength fluctuation in $k$ space, is
related to the detailed arrangement of the points on small scales. Our
study also shows how the exponents of the small-$k$ behavior of the 
structure factor
for these point processes encode properties of the tilings, and could
thus possibly be used as a method for classifying them. In a
related article by two of us \cite{ag-mj-cloudprocesses} the 
two-point correlation properties of point processes generated by 
replacing each particle, in a point process with known two point
properties, by a ``cloud''  of particles  are 
derived\footnote{Results in this case are derived in
\cite{ag-mj-cloudprocesses} under the assumption that the stochastic 
process describing the generation of the ``clouds'' and the initial
point process are independent. In the algorithm discussed here this
is not the case, as the points are ascribed to each tile in a
way which depends, in general, on the tile. For the particular case
of a Bravais lattice tiling, however, both calculations are valid
because of the equivalence of all tiles/points in such a  lattice.
Indeed, in this case, the different general formulae derived in the
present article and \cite{ag-mj-cloudprocesses}  give the same
result.}. One of us \cite{Uc06} numerically generated disordered
point distributions within a cubical box under periodic boundary
conditions with $\gamma > 4$. However, to
our knowledge, prior to this paper and \cite{ag-mj-cloudprocesses},
explicit analytical constructions of point processes 
with $\gamma > 4$ have not been given previously in the literature. 

It is instructive to recall qualitatively why tilings are a natural
starting point for the construction of superhomogeneous point
processes. A {\it tiling} or {\it tessellation} is a partition 
of Euclidean space $\mathbb{R}^d$  into
closed regions whose interiors are disjoint regions \cite{To02}.
Let us suppose we have a tiling of space by tiles
which are (i) of equal volume $\|T\|$, and (ii) bounded, 
with maximal length  $\Lambda$ in any direction. Let us now place
one point in each tile and consider the number fluctuations in the point
process so generated. If $N(R)$ is the number of points in a 
sphere of radius $R$, and of volume $V(R)$, it is simple
to see that 
\begin{equation}
\frac{V(R-\Lambda)}{\|T\|} \leq N(R) \leq \frac{V(R+\Lambda)}{\|T\|}\,.
\end{equation}
The lower bound is the minimal number of tiles which can overlap
the sphere of radius $R-\Lambda$ [and all such tiles must 
contribute a point to $N(R)$], the upper bound is the maximal
number of tiles which are fully enclosed in the sphere of radius 
$R+\Lambda$ [and only such tiles can contribute to $N(R)$].
For $R \rightarrow \infty$ we have therefore
\begin{equation}
|\Delta N(R)| \leq c R^{d-1}
\label{lim}
\end{equation}
where $c$ is a constant, and $\Delta N(R)= N(R) - \overline{N}(R)$
with $\overline{N}(R)=V(R)/\|T\|$. Averaging over configurations 
(or randomly placed centers for the spheres)  one anticipates that
{\em the slowest possible scaling of number fluctuations is}
\begin{equation}
\langle \Delta N^2(R) \rangle \propto R^{d-1}
\end{equation}
where $\langle...\rangle$ denotes the ensemble (or volume) average.
This behavior of the variance, proportional to the surface, is
a characteristic of superhomogeneous point processes. If there is 
appropriate long-range correlation in the tiling at arbitrarily 
large scale, the fluctuations could however, in principle, add coherently 
to give the more rapid growth up to
\begin{equation}
\langle \Delta N^2(R) \rangle \propto R^{2(d-1)}\,,
\end{equation}
which corresponds to the limit equality in Eq.~(\ref{lim}).
While in $d=1$ this still corresponds to surface 
fluctuations\footnote{Note that the case of $d=1$ is rather trivial
given our assumptions: the only equal volume tiling is the lattice.},
for any $d \geq 2$ it implies only the limiting small-$k$ behavior 
$S(k) \propto k^\gamma$ with $\gamma \ge -d+2$, which means that the 
point processes are not necessarily superhomogeneous for $d\ge 2$.
We will recover this result below, with the only difference that
the bound is found to be  $\gamma > -d+4$, which implies that
even long-range correlations between tiles give superhomogeneous processes 
for $d\leq 4$. The difference between this result and our naive estimate 
is simply due to the fact that below we constrain the particles to lie 
at the center of mass, rather than placing 
them randomly. In fact we will show here that by assigning more than
one point in an appropriately constrained manner to each tile, 
we can  increase these bounds on the exponents without limit,
and realize superhomogeneous processes with an arbitrary 
positive exponent, in any dimension.

\section{Point distributions from tilings: one point per tile}
\label{general formalism}

We consider in this section point processes generated by ascribing one
point to each tile. We first give a general analysis of the small-$k$
properties of the structure factor of the density fluctuations, and
derive how the leading behavior is determined by the properties of
the tiling. We then describe some specific explicit constructions
which illustrate the result.
    
\subsection{Density fluctuations and long-wavelength limit}
\label{formula}

We start from a generic (regular or irregular) 
tiling\footnote{A regular tiling 
is periodic in space. An irregular tiling is aperiodic in space,
including quasiperiodic as well as disordered tilings. A congruent
tiling consists of identical tiles.} 
of $d$-dimensional 
Euclidean space $\mathbb{R}^d$ into {\it equal volume tiles}, which 
we denote $T_i$. We consider the point distribution generated by 
ascribing  one point to each tile, and placing it at position {\bf x}$^i$, 
which coincides with the center of mass of the tile $T_i$, i.e.,
\begin{equation}
\label{def-cm}
{\bf x}_i= \frac{1}{\| T \|} \int_{T_i} d^dx \,{\bf x}
\end{equation}
where $\| T \|$ is the volume of the tiles. The density fluctuation 
field is thus
\begin{equation}
\label{density}
\delta n ({\bf x}) = 
\sum_i \delta^{(d)} \left( {\bf x} - {\bf x}_i \right)-n_0
\end{equation}
where $n_0$ is the mean number density in the infinite-volume 
limit\footnote{Since all particles 
have the same mass no distinction need be made between the mass and number 
density fluctuations. For the case of a single particle per tile
$n_0 =1/\| T \|$.}, and $\delta^{(d)}({\bf x})$ is the Dirac delta 
function in $d$ dimensions.
The structure factor (SF) is defined as 
\begin{equation}
S({\bf k}) = \lim_{V \rightarrow \infty} 
\frac{|\tilde{\delta n} ({\bf k}; V)|^2}{n_0 V}= 1+ n_0 {\tilde h}({\bf k})
\end{equation}
where $V$ is the system volume,
\begin{equation}
\tilde{\delta n} ({\bf k}; V) = \int_V d^d x e^{-i {\bf k}\cdot{\bf x}}
\delta n ({\bf x})\,,
\end{equation}
and 
\begin{equation}
{\tilde h}({\bf k})=\int_{\mathbb{R}^d} d^d x e^{-i {\bf k}\cdot{\bf x}} 
h({\bf r})
\end{equation}
is the infinite-space Fourier transform of the total pair correlation
function $h({\bf r})$, which vanishes for disordered systems when the
distance $r\equiv |{\bf r}|$ tends to infinity \cite{To03,Ga04}.

For our point process it follows directly that
\begin{equation}
\tilde{\delta n} ({\bf k}; V) = 
\sum_i e^{-i {\bf k}\cdot{\bf x}_i} \left[1 -\tilde{W_i}({\bf k})\right] 
\end{equation}
where the sum runs over the points enclosed in the 
volume $V$, and $\tilde{W_i}({\bf k})$ is the
normalized characteristic function of the tile $T_i$, given by   
\begin{equation}
\label{characteristic}
\tilde{W_i} ({\bf k}) = \frac{1}{\| T \|}
\int_{T_i(0)} d^d x e^{-i {\bf k}\cdot{\bf x}}\,,
\end{equation}
where $T_i(0)$ denotes that the center of mass of the tile
has been taken as the origin of axes. If we assume that $\tilde{W_i}$
is an analytic function at ${\bf k}=0$,
we can expand it in Taylor series, to obtain
\begin{equation}
\tilde{W_i} ({\bf k}) = 1 + \sum_{m=2}^{\infty} \frac{(-i)^m}
{m!} k_{\alpha_1}...k_{\alpha_m} M_{\alpha_1...\alpha_m} (i)
\label{eq12}
\end{equation}
where 
\begin{equation}
M_{\alpha_1...\alpha_m} (i) = \frac{1}{\| T \|} \int_{T_i(0)} d^d x \,
x_{\alpha_1}...x_{\alpha_m}
\label{mass-moments}
\end{equation}
is a (fully symmetric) tensor of rank $m$  corresponding to
the $m$-th moment of the mass distribution of the tile
$T_i$ (normalized by the volume/total mass) and  
$\alpha_j=1,\ldots,d$ are indices for the Cartesian components. 
Note that we have used 
Eq.~(\ref{def-cm}), which makes the linear term in the expansion
(\ref{eq12}) (corresponding to the dipole moment) vanish. 
The assumption of analyticity corresponds to
the requirement that all these moments are finite. This is true
in particular if the tiles are of finite extent. We will discuss briefly in
our conclusions the possibility of relaxing this assumption.

Using these expressions in the definition of $S({\bf k})$ we now
obtain
\begin{eqnarray}
\label{Sk-calI}
&&S({\bf k}) =\\ 
&&\sum_{n=2}^{\infty} \sum_{m=2}^{\infty}
\frac{(-i)^{m} (i)^{n}}{m!\, n!} 
k_{\alpha_1}...k_{\alpha_n} k_{\beta_1}...k_{\beta_m} 
{\cal I}_{\alpha_1...\alpha_n \beta_1...\beta_m} ({\bf k})\nonumber
\end{eqnarray}
where 
\begin{eqnarray}
\label{defn-calI}
{\cal I}_{\alpha_1...\alpha_n \beta_1...\beta_m} ({\bf k})
=\lim_{V \rightarrow \infty} \frac{1}{N} \sum_i \sum_j 
e^{-i {\bf k} \cdot ({\bf x}_i - {\bf x}_j)} \nonumber \\
\times M_{\alpha_1...\alpha_n} (i) M_{\beta_1...\beta_m} (j)
\end{eqnarray}
where the sums run over the $N$ particles contained in the volume $V$. 
It is straightforward to verify that the 
coefficient of the leading term in $k$ (at order $k^4$) 
is non-negative, and that the coefficients of all powers 
of $k$ are real. Indeed the SF $S({\bf k})$ is
by definition a real non-negative quantity, and Eq.~(\ref{Sk-calI}) is 
just the specific form of its Taylor expansion around ${\bf k}=0$ 
for the particular class of distributions we are considering. 

It is convenient to rewrite the latter expression as
\begin{eqnarray}
{\cal I}_{\alpha_1...\alpha_n \beta_1...\beta_m} ({\bf k})
=\lim_{V \rightarrow \infty} \frac{1}{n_0 V} 
\int d^d x\, d^d y\, e^{-i {\bf k} \cdot ({\bf x} - {\bf y})} 
 \nonumber \\
{\cal M}_{\alpha_1...\alpha_n} ({\bf x}) {\cal M}_{\beta_1...\beta_m} ({\bf y})
\end{eqnarray}
where 
\begin{equation}
\label{tensor-field definition}
{\cal M}_{\alpha_1...\alpha_n} ({\bf x}) 
= \sum_{i} \delta^{(d)}({\bf x}-{\bf x}_i)
M_{\alpha_1...\alpha_n} (i)\,.
\end{equation}
The distribution ${\cal M} ({\bf x})$ can be viewed as a weighted 
particle density. The weight associated with each particle is 
the appropriate component of the mass moment of the tile to
which the particle belongs.

Up to now we have considered implicitly a single
particle placed deterministically in each tile. We now consider 
averaging over an appropriately defined ensemble of
such tilings\footnote{For a deterministic tiling, e.g. the cells of a
regular lattice, or the pinwheel tiling in two dimensions discussed
below, this average can be defined by the set of configurations
generated by applying an arbitrary rigid translation to a given
configuration.}.  If the tiling is statistically translationally invariant,
we have that
\begin{equation}
\langle {\cal M}_{\alpha_1...\alpha_n} ({\bf x}) 
{\cal M}_{\beta_1...\beta_m} ({\bf y}) \rangle 
\equiv g^{n,m} ({\bf x} - {\bf y}) \,,
\end{equation}
where $\langle...\rangle$ denotes the ensemble average. We have adopted 
here for the correlation function $g^{n,m}({\bf x})$ the tensorial notation in
which the indices are left implicit. In this notation we can write our
result for the SF $S({\bf k})$ as  
\begin{equation}
\label{sk-tensor}
S({\bf k}) = \frac{1}{n_0} \sum_{n=2}^{\infty} \sum_{m=2}^{\infty}
\frac{(-i)^{m} (i)^{n}}{m!\, n!} 
{\bf k}^n \cdot \tilde{g}^{n,m} ({\bf k}) \cdot {\bf k}^m  
\end{equation}
where 
$\tilde{g}^{n,m} ({\bf k})$ is the Fourier transform of $g^{n,m} ({\bf x})$
defined as 
\begin{equation}
\tilde{g}^{n,m} ({\bf k})
=\int d^d x\, e^{-i {\bf k} \cdot {\bf x}} 
g^{n,m} ({\bf x})\,,
\end{equation}
and ${\bf k}^n$ denotes a tensor of order $n$, given by the tensor
product of $n$ vectors ${\bf k}$, i.e., 
\begin{equation}
{\bf k}^n_{\alpha_1...\alpha_n} 
\equiv [{\bf k}\otimes {\bf k}...\otimes{\bf k}]_{\alpha_1...\alpha_n}
= k_{\alpha_1} k_{\alpha_2} ..k_{\alpha_n}\,.
\end{equation}
The symbol $\cdot$ in Eq.~(\ref{sk-tensor}) denotes the contraction of
the corresponding tensor indices. If the ensemble is also statistically 
isotropic the product 
${\bf k}^n \cdot \tilde{g}^{n,m} ({\bf k}) \cdot {\bf k}^m$, and thus
$S({\bf k})$, is a function of $k=|{\bf k}|$ only. 

The behavior of $S({\bf k})$ at small $k$ is thus manifestly
determined by that of $\tilde{g}^{n,m} ({\bf
k})$ in this limit. These quantities are
in fact the (tensor) SFs associated to the 
discrete stochastic field defined by Eq.~(\ref{tensor-field definition}), 
the two point correlation function of 
which is $g^{n,m} ({\bf x})$. 
They thus encode information about the tiling, and more 
specifically about the correlation properties of the second and higher
moments of the tiles. We restrict ourselves to the case 
that all these moments are finite, and strictly bounded
(which also ensures, as noted above, the validity of the expansion 
of the characteristic function $\tilde{W_i}({\bf k})$ we have performed).
As noted above, each component of the tensor stochastic 
field ${\cal M}^n ({\bf x})$ defined 
in Eq.~(\ref{tensor-field definition}) above
is then a discrete stochastic process in which the points are 
located at the same positions as in the point process we are studying,
but have ``masses''  given by the corresponding component of the tensor 
$M^n(i)$ (or rather ``charges'' as they are not
strictly positive) which are bounded (above and below).
Just as for a generic stochastic point process such
a discrete (or indeed continuous) process can be classified
into three categories according to the small-$k$ behavior
of the $\tilde{g}^{n,m} ({\bf k})$:

\begin{enumerate}
\item $\tilde{g}^{n,m} ({\bf k}=0) = {\rm const.} > 0$:
this means that the correlation functions  $g^{n,m} ({\bf x})$ 
of the higher order moments are integrable at large $x$, 
and the corresponding integral is equal to a positive 
constant, i.e., the higher moments of the tiles have short-range
correlations dominated by the positive contributions. 
For the generated point process  we have then
the leading behavior $S({\bf k} \rightarrow 0) \propto k^4$.

\item   $\tilde{g}^{n,m} ({\bf k}=0) = 0$: the integral of the
correlation functions $g^{n,m} ({\bf x})$ converges to zero, i.e., the
shapes and orientations of the tiles have themselves superhomogeneous
properties (i.e. in which the positive and negative correlations 
balance exactly in the integral). In this case we will 
obtain a leading behavior
$S({\bf k} \rightarrow 0) \propto k^\gamma$ with $\gamma > 4$ [and
with a value depending on the leading behavior of the
$\tilde{g}^{n,m} ({\bf k})$ at small-$k$].

\item  $\tilde{g}^{n,m} ({\bf k}=0) = \infty$, with 
$\tilde{g}^{n,m} ({\bf k} \rightarrow 0) \propto k^\alpha$
and $-d<\alpha < 0$. In this case, in which the correlation 
functions $g^{n,m} ({\bf x})$ are non-integrable, i.e., 
the higher moments of the tiles have themselves long-range 
correlations, we can obtain a leading behavior for our
point process $S({\bf k} \rightarrow 0) \propto k^\gamma$ 
with $4-d < \gamma < 4$.

\end{enumerate}

Several remarks on this result are important. Firstly, for 
simplicity we have assumed above that the 
$\tilde{g}^{n,m} ({\bf k})$ are in the same class for all
$m$ and $n$. This is, of course, {\it a priori}, not necessarily
the case. In the more general case that the different  
$\tilde{g}^{n,m} ({\bf k})$ are in different classes, the 
determination using Eq.~(\ref{sk-tensor}) of the exponent 
$\gamma$ of the leading small-$k$ behavior of $S({\bf k})$ 
is nevertheless straightforward. Furthermore, it is simple to verify
that the bounds we have given on this exponent remain valid. 

Secondly, we have assumed that the $\tilde{g}^{n,m} ({\bf k})$ 
obey the condition
\begin{equation}
\label{convergence}
\lim_{k \rightarrow 0} k^d \tilde{g}^{n,m} ({\bf k}) =0\,.
\end{equation}
This assumption corresponds to the requirement that the discrete 
processes defined by Eq.~(\ref{tensor-field definition})
have well-defined mean values, i.e, the normalized 
fluctuations (e.g. integrated in a sphere) of the moments 
of the tiles converge to zero in the infinite-volume limit. 
While this seems a very weak assumption, it is not a 
priori true of all tilings. 

\subsection{Explicit constructions}
\label{explicit:one point per tile}

We now give various explicit constructions to illustrate the above
results. 

\subsubsection{Regular lattice tilings}

Consider first the tiling given by the Voronoi cells of any Bravais lattice.
In a Bravais lattice, the space $\Re^d$ can be geometrically divided into identical
regions $F$ called {\it fundamental cells}, each of which contains just
one point of the lattice.  A {\em Voronoi cell} associated
with a point at ${\bf r}$ in a point distribution is
defined to be the region of space nearer
to the point at ${\bf r}$ than to any other point.
Since all Voronoi cells or tiles of a Bravais lattice
have the same shape and orientation, the mass moments 
$M^{n} (i)$ in Eq.~(\ref{mass-moments}), calculated with respect to
the center of mass, are identical for all tiles, i.e., $M^{n} (i)$
does not depend on $i$. The discrete processes specified by 
Eq.~(\ref{tensor-field definition}) are then simply, up to
a constant, equal to the density field of the lattice, and thus
it follows that all the correlation functions $g^{n,m} ({\bf x})$ 
are proportional to the two point correlation function of the 
original lattice. Thus $\tilde{g}^{n,m} ({\bf k})$, and 
also $S({\bf k})=0$, is zero in some finite region around 
${\bf k}=0$\footnote{More
precisely $S({\bf k})=0$ at all ${\bf k}$ different from non-zero
reciprocal lattice vectors.}. This result is in fact evident:  
the point process generated by placing points at the center of
mass of every cell is of course simply the lattice itself.
In this case, of course, neither the tiling nor the point process 
are statistically isotropic (where the statistical average is taken 
over lattices rigidly translated within the elementary lattice cell)

The same result can evidently be generalized to any tiling 
of equal volume cells constructed from a periodic point pattern.

\subsubsection{Congruent rotationally invariant tilings}
\label{Congruent rotationally invariant tilings-1}

We consider next deterministic congruent tilings with the additional
property of rotational invariance, i.e., in which all orientations of
the identical tiles are equiprobable.  Known examples are the
pinwheel tiling \cite{pinwheel} in $d=2$ (see Fig. 1) and the quaquaversal tiling
\cite{quaquaversal} in $d=3$. In these cases each tile $T_i$ can be
characterized solely by its center of mass ${\bf x}_i$ and by a
matrix $R(i) \in 0(d)$, the latter giving the orientation with 
respect to some arbitrary chosen orientation. 

\begin{figure}[htbp]
\begin{center}
\includegraphics[width=9cm]{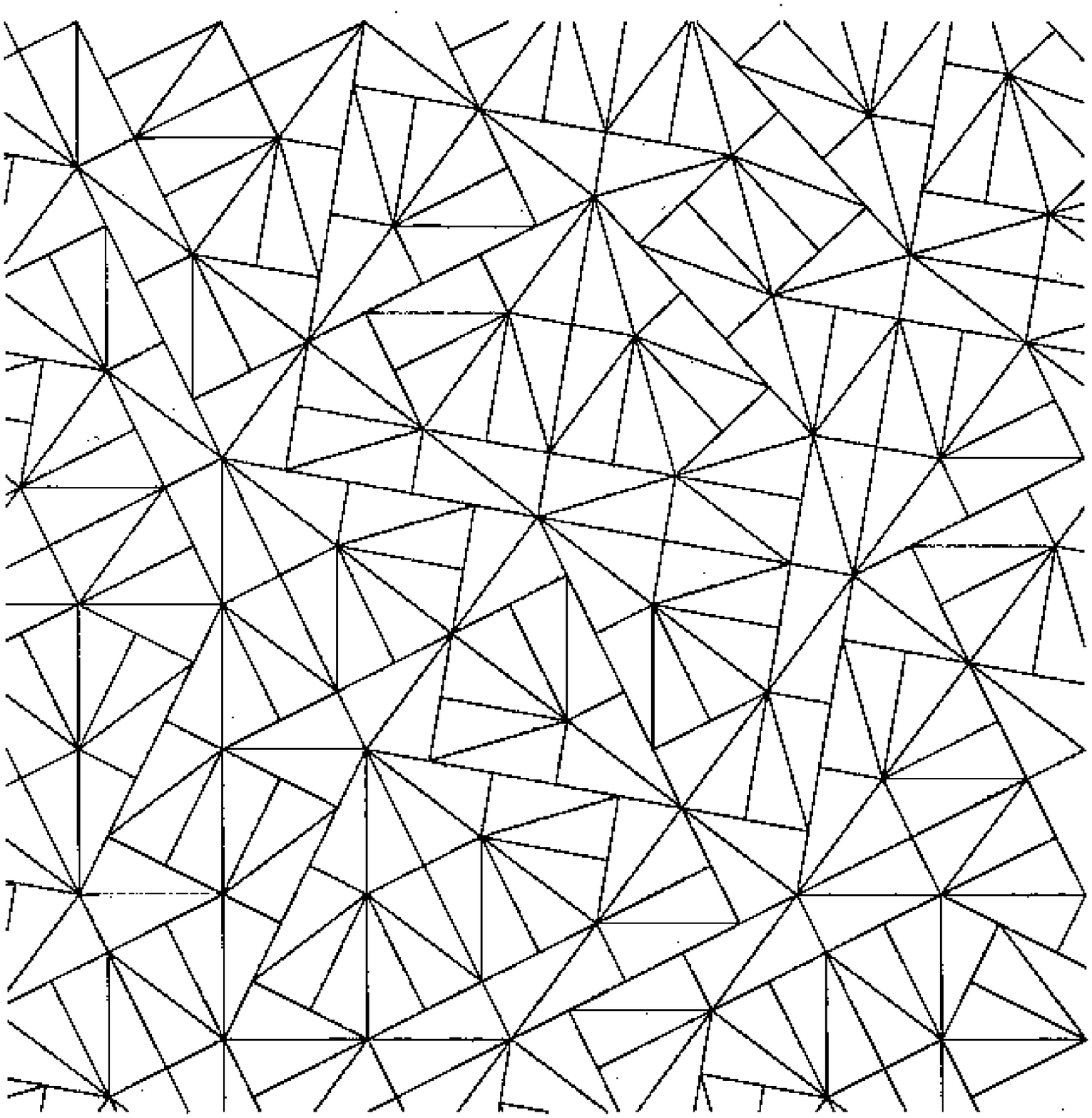}
\caption{Portion of a pinwheel tiling. The prototile of the pinwheel tiling
is a right triangle with sides of length one, two, and $\sqrt{5}$. The
tiling is produced by performing certain ``decomposition''
and ``inflation'' operations on the prototile. In the first step, the prototile
is subdivided into five copies of itself and then these new triangles
are expanded to  the size of the original triangle. These decomposition
and inflation operations are repeated {\it ad infinitum} until the triangles
completely cover the plane.
\label{fig1}}
\end{center}
\end{figure}

We can then write, in tensorial notation, 
\begin{equation}
\label{tensor-field-congruent}
{\cal M}^{n} ({\bf x}) 
= \sum_{i} \delta^{(d)}({\bf x}-{\bf x}_i)
R^{n}(i) \cdot {\cal \tilde{M}}^{n}
\end{equation}
where $R^n(i)$ is the tensor product, giving
a tensor of rank $2n$ of which $n$ indices are
contracted with the corresponding
moment ${\cal \tilde{M}}^{n}$ of a tile with the reference 
orientation, i.e., 
\begin{equation}
\label{tensor-notation}
[R(i)^{n} \cdot {\cal \tilde{M}}^{n}]_{\alpha_1...\alpha_n} 
\equiv 
R_{\alpha_1 \beta_1 }(i)... R_{\alpha_n \beta_n }(i)
{\cal \tilde{M}}^{n}_{\beta_1...\beta_n} 
\end{equation}
where, as everywhere above, the sums over the indices which
appear twice are implicit. The correlation functions 
$g^{n,m} ({\bf x})$ are thus
direct measures of the correlation of the {\it orientations}
of tiles with center of mass separated by ${\bf x}$.

The case of the quaquaversal tiling has been studied 
numerically in Ref.~\cite{Ha06}, and an approximate small-$k$ 
behavior $S({\bf k}) \propto k^4$ found. In Ref.~\cite{white_WDM}
it is noted, however, that the numerical results agree better 
with $S({\bf k}) \propto k^\gamma$ and $\gamma \approx 3.4$.
While the former behavior would correspond, as discussed above,
to a short-range correlation of the orientation of the tiles,
the latter would instead correspond to a weak long-range 
correlation (with a correlation function characterizing the
orientations decaying with distance $r$ as $\sim r^{-2.4}$).
Further, superimposed on this power-law behavior there
are residual peaks at certain wavenumbers, with a spacing 
which appears to be consistent \cite{Ha06} with the hierarchical
nature of the tiling \cite{radin_JSP1999}.
As we have discussed the small-$k$ behavior of the constructed
point process thus probes the correlation properties of the underlying
tiling.

\subsubsection{Random Binary Rectangular (RBR) tiling}
\label{Random Binary Rectangular (RBR) tiling}

It is instructive to illustrate our result with a non-trivial
example which, albeit not statistically isotropic, allows us to 
calculate exactly the SF of a point process 
constructed by the algorithm we have described.
The example we now give is of a stochastic congruent tiling.
For simplicity we work in $d=2$, but a generalization to any 
$d$ is straightforward. 

\begin{figure}[htbp]
\begin{center}
\includegraphics[height=10cm, width=9cm]{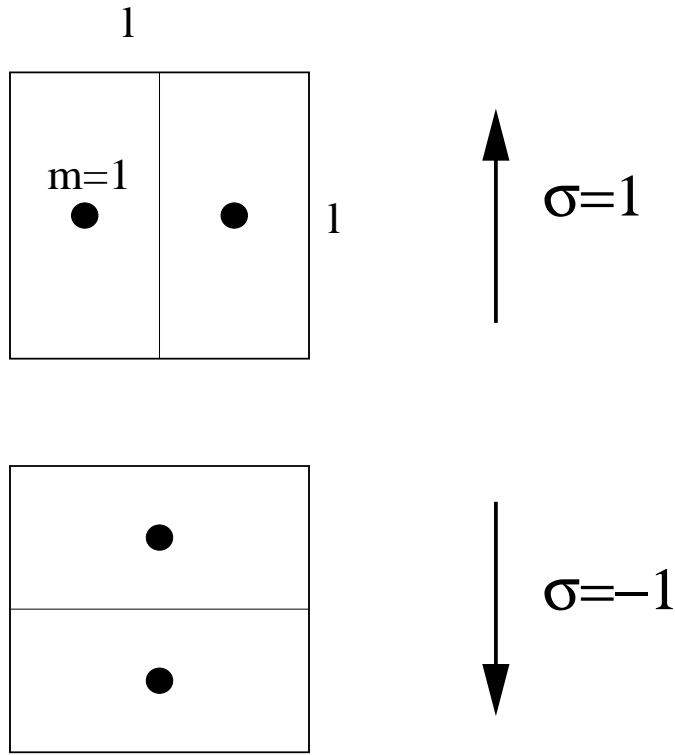}
\caption{Elementary binary rectangular tiling, with a unit mass
particle in the center of mass of each rectangle, and its description
in terms of an Ising-like spin variable.
\label{fig2}}
\end{center}
\end{figure}

We generate the tiling as follows. We start from a regular tiling 
of the plane with congruent squares. We then divide each square tile in half,
defining two identical rectangular sub-tiles, as shown in Fig.~\ref{fig2}.
The choice of the orientation of each tile is given by a stochastic
process, which can be cast as the value of a simple up-down
spin variable. The density of the point process generated using the 
algorithm analyzed in the previous section, in which a point
is placed at the center of mass of each tile, can then be written 
\begin{eqnarray}
n(\bx) = \sum_{\bR}\sum_{n=1}^2\delta 
\left[x-R_x-\frac{1+\s_\bR}{2}(-1)^n\right] 
\nonumber \\
\times \delta\left[y-R_y-\frac{1-\s_\bR}{2}(-1)^n\right]
\end{eqnarray}
where $\bR\equiv (R_x,R_y)$ are the lattice sites of the underlying
square lattice placed at the center of each square cell, and
$\s_\bR=\pm 1$ is the spin variable specifying the orientation of the
two elementary rectangular tiles at the lattice site $\bR$ as in
Fig.~\ref{fig2}. We assume that the lattice spacing of the underlying
square lattice is $l$. Consequently the volume of the elementary
rectangular tile is $l^2/2$ and therefore the average number density
of the point process is $n_0=2/l^2$.  
It is simple to show that the Fourier transform (FT) 
of $[n(\bx)-n_0]$ is
\begin{eqnarray}
&&\tilde{\delta n}(\bk;V)=2\sum_{\bR}e^{-i\bk\cdot\bR} 
\left[\cos\left(k_xl\frac{1+\s_\bR}{8}+k_yl\frac{1-\s_\bR}{8}\right)\right.
\nonumber\\
&& 
\left.-\frac{\sin (k_xl/2)}{k_xl/2}\frac{\sin(k_yl/
2)}{k_yl/2} \right]\,.
\label{eq1}
\end{eqnarray}
To calculate the SF averaged over the ensemble of possible 
configurations of the binary tiles we assume that $\left<\s_\bR\right>=0$ 
(i.e. both orientations of the binary tiles are equiprobable), 
and write $\chi(\bR)=\left<\s_{\bR_0}\s_{\bR_0+\bR}\right>$ 
using the lattice statistical translational invariance.
Moreover, since $\s_\bR=\pm 1$, we have that 
\begin{eqnarray}
\label{eq3}
&& \cos\left(k_xl\frac{1+\s_\bR}{8}+k_yl\frac{1-\s_\bR}{8}\right)\nonumber \\
&&=\frac{1+\s_\bR}{2}\cos\left({k_xl\over 4}\right)+\frac{1-\s_\bR}{2}
\cos\left({k_yl\over 4}\right)\,.
\end{eqnarray}
It is then straightforward to obtain the following exact expression
for the SF: 
\begin{eqnarray}
S(\bk)&=&\frac{[\cos (k_xl/4)-\cos (k_yl/4)]^2}{2}\sum_\bR
e^{-i\bk\cdot\bR}\chi(\bR)\nonumber\\
&+&S_{ML}(\bk)
\label{eq4}
\end{eqnarray}
where 
\begin{equation}
S_{ML}(\bk)=A(\bk)\sum_\bR e^{-i\bk\cdot\bR}=\pi^2\sum_{\bH\ne
0}A(\bH)\delta(\bk-\bH)
\end{equation}
is a ``modulated lattice" SF which is different from
zero only at the non-zero reciprocal lattice vectors $\bH$, and
\bea
A(\bk)&=&\frac{1}{2}\left[\cos\left({k_xl\over 4}\right)+\cos\left(
{k_yl\over 4}\right)\right.\nonumber\\
&-&\left.2\frac{\sin(k_xl/2)}{k_xl/2}
\frac{\sin(k_yl/2)}{k_yl/2}\right]^2\,.
\label{A-k}
\eea
At small-$k$ only the first term of Eq.~(\ref{eq4}) contributes, and
expanding the factor outside the sum we thus obtain a leading 
small-$k$ behavior
\begin{equation}
S(\bk)\simeq \left({l\over 4}\right)^4(k_x^2- k_y^2)^2 \tilde \chi(\bk)
\end{equation}
where 
\begin{equation}
\tilde\chi(\bk)= \lim_{N \rightarrow \infty} \frac{1}{N} \sum_\bR
e^{-i\bk\cdot\bR}\chi(\bR)\,.
\end{equation}
This exact result is of course a special case of the general analysis
given above, in which the moments characterizing the tiles are
particularly simple as there are only two orientations. All the
two-point properties of the point process are then contained in the
single correlation function $\chi(\bR)$ of these orientations for
pairs of tiles with centers separated by $\bR$. As $\tilde \chi(\bk)$
is a power spectrum of a stochastic process with a well defined mean,
at small $k$ we have $\tilde\chi(k)\sim k^b$ with $b>-d\equiv-2$. If
$\tilde\chi(0)=c>0$, i.e., in the case in which the orientations of the
tiles are short-range correlated, then
$S(\bk)=O(k^4)$ at small $k$. If instead
$\tilde\chi(k\rightarrow\infty)=\infty$ there are long-range positive
correlations in the orientations of the tiles which induce a slower decay of
the fluctuations in the associated point process at large scales.
Finally if $\tilde\chi(0)=0$ the stochastic spin process is itself
superhomogeneous, with a balance between positive and negative
correlations creating a sort of ``stochastic order" in the spin
configuration.  In this case $S(\bk)$ vanishes faster than $k^4$ at
small $k$. In Appendix \ref{appendix} we describe explicitly an
algorithm for generating such spin configurations.

\section{Point distributions from tilings: $n>1$ points per tile}
\label{general formalism2}

The algorithm described in Sect.~\ref{Random Binary Rectangular (RBR)
tiling} can in fact be thought of in a different way to that in which 
we have presented it: one can consider
it instead as a direct assignment of two points to the tiles of the
original square lattice, without any construction of an intermediate
SBR tiling. The pair of particles then have two possible orientations,
which are chosen stochastically. Note that in each case the center of
mass of the particles is located at the center of the square cell. Our
results above shows that if this stochastic process is short-ranged
correlated we obtain a small-$k$ behavior
$S(k) \propto k^4$, while with a single point at the center of mass of
the lattice cell we recovered (evidently) the lattice. We now consider
quite generally what small-$k$ behavior of $S(k)$ of a point process
we can obtain by ascribing more than one point per tile in a generic
tiling with equal volume tiles.

\subsection{Density fluctuations and long-wavelength limit}
\label{formula2}

We ascribe $p$ points to each tile, denoting their positions
by
\begin{equation}
{\bf x}^{i,\ell}={\bf x}^{i} + {\bf u^{i,\ell}}
\end{equation}
with $\ell=1...p$ and ${\bf x}^{i}$ is, as above, the center of
mass of the tile $T_i$ (so that ${\bf u^{i,\ell}}$ is the 
position relative to the center of mass). 
We assume further that
the center of mass of the points coincides with that of the tile,
i.e., 
\begin{equation}
\sum_{\ell=1}^p {\bf u^{i,\ell}}=0\,.
\end{equation}
Following the same steps as in Sect.~\ref{general formalism}, we arrive at
\begin{equation}
\tilde{\delta n} ({\bf k}; V) = 
\sum_i e^{-i {\bf k}\cdot{\bf x}^i}
\left[\frac{1}{p} \sum_{\ell=1}^{p} e^{-i {\bf k}\cdot{\bf u}^{i,\ell}}
-\tilde{W_i}({\bf k})\right] 
\end{equation}
where $\tilde{W_i}({\bf k})$ is precisely the same normalized characteristic 
function of the tile as defined in Eq.~(\ref{characteristic}) [and
$1/p$ is the fraction of the mass of the tile ascribed to each particle].
Expanding in Taylor series we obtain
\begin{equation}
\label{expansion-p-points}
\frac{1}{p} \sum_{\ell=1}^{p} e^{-i {\bf k}\cdot{\bf u}^{i,\ell}}
-W_i({\bf k})= \sum_{m=2}^{\infty} \frac{(-i)^m}
{m!} k_{\alpha_1}...k_{\alpha_m} M_{\alpha_1...\alpha_m} (i)
\end{equation}
where now
\begin{equation}
\label{difference}
M_{\alpha_1...\alpha_m} (i) = 
\frac{1}{p} \sum_{\ell=1}^{p} 
u^{i,\ell}_{\alpha_1}...u^{i,\ell}_{\alpha_m}
-\frac{1}{\| T \|} \int_{T_i(0)} d^d x \, x_{\alpha_1}...x_{\alpha_m}
\end{equation}
is the totally symmetric rank $m$ tensor corresponding to
the {\it difference} of the $m$-th moments of the mass distribution
of the points associated with the tile $T_i$ and that of the 
tile itself. As in the derivation with one point, we have assumed 
the analyticity of the quantity we expanded. This means that we
require that all the moments in Eq.~(\ref{difference}) are finite, 
which is true in particular if all the tiles are of finite extent 
and the lengths of the vectors ${\bf u}^{i,\ell}$ are bounded. 

All the expressions, and notably the result for $S({\bf k})$ in
Eq.~(\ref{sk-tensor}), derived in the case with one point per tile,
are thus valid. The only difference is that the tensors $M^n(i)$ 
[and correspondingly ${\cal M}^n(\bx)$] are now given 
by Eq.~(\ref{difference}). 
The small-$k$ properties thus 
depend, assuming statistical translational invariance, on those of 
the FT of the correlation functions of the 
{\it differences} of the moments of the discrete mass distribution 
of the points ascribed to each tile and that of the continuous mass
distribution represented by the tile itself.  

The most important implication of this result is the following:
the coefficients in the small-$k$ expansion of 
Eq.~(\ref{expansion-p-points}) are now proportional
to a {\it difference} of two quantities in Eq.~(\ref{difference}). 
Any given coefficient will vanish identically  
if our assignment of the $p$ points satisfies the
constraints
\begin{equation}
\label{constraints}
\frac{1}{p} \sum_{\ell=1}^{p} 
u^{i,\ell}_{\alpha_1}...u^{i,\ell}_{\alpha_m}
b=\frac{1}{\| T \|} \int_{T_i(0)} d^d x \, x_{\alpha_1}...x_{\alpha_m}
\end{equation}
{\it in each tile}, i.e., if {\it the (tensorial) moments of the mass
distribution of the points are equal to those of the tile in which
they are placed}.  With a sufficiently large number of points per
tile one can evidently make any desired finite number of terms vanish
in the expansion of $S(\bk)$, i.e., construct a stochastic point mass
distribution for each tile which has all moments up to a certain order
equal to those of the continuous mass distribution of the tile. 

This procedure allows one
to obtain an arbitrarily large exponent $\gamma$ in the small-$k$
behavior of the SF of the point process.

\subsection{Explicit constructions}

We again illustrate these results with some explicit examples.

\subsubsection{Regular lattice tiling}

In a Bravais lattice, as we have discussed above, the tensorial
moments of all {\it tiles} are equal so that the second term in
Eq.~(\ref{difference}) does not contribute to $S({\bf k})$ in a finite
region around ${\bf k}=0$. This can most easily be seen by using 
Eq.~(\ref{difference}) directly in the expressions 
Eqs.~(\ref{Sk-calI}) and (\ref{defn-calI}) for $S({\bf k})$:
the $i$-independent second term in each component of the tensor
$M^n(i)$, when summed over $i$, gives a delta function proportional 
to the SF of the lattice.  If we have more than 
one point per tile (i.e., $p \geq 2$) we may have, however, 
an $i$-dependent contribution from the first term in 
Eq.~(\ref{difference}), i.e., from the moments of the mass
distribution constituted by the points assigned to the single
tile, which are not constrained (beyond the dipole moment).  
If we allow  the distribution of these points to vary stochastically from 
cell to cell, we will generically have a non-zero contribution 
for the SF (ensemble averaged over the stochastic process) 
at all $k$, i.e., we will have a continuous SF\footnote{If, on the other hand, the points are placed
in the same way with respect to the center of mass of each tile, $S({\bf k})$
is again zero in the same region. The remaining non-zero piece has a
modulated delta-function structure which can be easily calculated. Indeed
the point distribution so generated is in this case again a periodic
lattice, i.e., the initial Bravais lattice with basis of $p$ points 
per cell.}.

As a simple example let us consider first the RBR algorithm analyzed
above, cast as the ascription of two points to each cell of a simple
cubic lattice, but now allowing the pair of particles ascribed
to each cubic cell have a random orientation, i.e., we take 
two points in each lattice cell with coordinates
\begin{equation}
\label{define-SSL}
\bx_1 (\bR)= \bR + {\bf u} (\bR) \,,\qquad
{\bx}_2 (\bR)= \bR - {\bf u} (\bR) 
 \end{equation}
where $\bR$ is the generic lattice site (which is the
center of mass of the corresponding tile), and the 
vectors ${\bf u}(\bR)$
are generated by a stochastic process. We assume further that 
the vectors ${\bf u}(\bR)$  in different lattice cells are 
uncorrelated. The ensemble is thus fully specified by the one 
point probability distribution function $p({\bf u})$.
We call this stochastic point process the {\it split shuffled lattice} 
\footnote{In 
the context of the context of causality bounds on fluctuations
in cosmology, this construction has been studied
in Ref.~\cite{wandelt}.}, as it is
a generalization of the {\it shuffled lattice}
discussed in Ref.~\cite{glasslike} (see also Refs.~\cite{andrea, book}),
in which one point is randomly displaced off a perfect 
lattice. 

The leading small-$k$ behavior of $S(\bk)$ in this case may
be found easily by taking the ensemble average of the leading term 
in Eq.~(\ref{Sk-calI}):
\begin{equation}
\label{Sk-calI-leading}
S({\bf k}) = 
\frac{1}{4} 
k_{\alpha}k_{\beta} k_{\gamma}k_{\delta} 
\langle {\cal I}_{\alpha \beta \gamma \delta} ({\bf k})\rangle
\end{equation}
where, using Eq.~(\ref{difference}) in Eq.~(\ref{defn-calI}),  
we have 
\begin{eqnarray}
\label{defn-calI-leading-lattice}
\langle{\cal I}_{\alpha \beta \gamma \delta} ({\bf k}) \rangle
=\lim_{V \rightarrow \infty} \frac{1}{N} \sum_{\bR} \sum_{\bR^\prime} 
e^{-i {\bf k} \cdot ({\bR} - {\bR^\prime})} \nonumber \\
\times \langle u_{\alpha}(\bR) u_{\beta} (\bR)
u_{\gamma} (\bR^\prime) u_{\delta} (\bR^\prime) \rangle \,.
\end{eqnarray}
Since the vectors $\bu (\bR)$ are, by assumption, uncorrelated 
at different sites, we have
\begin{equation}
\langle u_{\alpha}(\bR) u_{\beta} (\bR)
u_{\gamma} (\bR^\prime) u_{\delta} (\bR^\prime) \rangle 
= \langle u_{\alpha} u_{\beta} \rangle
\langle u_{\gamma} u_{\delta} \rangle \,.
\end{equation}
for $\bR \neq \bR^\prime$. Using the fact, again, that the sum
$\frac{1}{N} \sum_{\bR} \sum_{\bR^\prime} 
e^{-i {\bf k} \cdot ({\bR} - {\bR^\prime})}$ is proportional
to the SF of the original
lattice, which is zero around ${\bf k}=0$, we can then write the 
leading small-$k$ behavior as
\begin{equation}
\label{sk-SSL}
S(\bk) = \frac{1}{4} \left[ \langle (\bk \cdot \bu )^4 \rangle
- \langle (\bk \cdot \bu )^2 \rangle ^2 \right] \,.
\end{equation}
If we choose a probability distribution which is isotropic
in ${\bu }$, i.e., $p({\bu }) \equiv p(u)$ 
(with $u=|{\bu }|)$, we then have 
\begin{equation}
\label{sk-SSL-result}
S(\bk)= \frac{k^4}{4d^2} \left[ C(d) \langle u^4 \rangle
- \langle u^2 \rangle ^2 \right]\,.
\end{equation}
where $C(d)\geq 1$ is a constant [$C(1)=1$, $C(3)=9/5$].  
We thus obtain the same exponent $\gamma$ as in the RBR tiling
model, but now the fluctuations at small $k$ are isotropic at leading
order because of the isotropy in the attribution of the displacement
vectors. Note that we have assumed here, as we do throughout this
paper, that the moments of the mass distribution in each tile are
finite, which requires here manifestly in Eq.~(\ref{sk-SSL-result})
the finiteness of at least the first four moments of $p(u)$. It is
simple, however, to generalize this kind of model to the case when
moments of order lower than the fourth diverge, just as done for
the shuffled lattice in Ref.~\cite{andrea} (see also
\cite{gabrielli_etal_04}). In this case one can 
obtain any small-$k$ behavior ($0 <\gamma \leq 4$).  

This kind of algorithm can easily be generalized, conserving a
sufficiently large number of moments in such a way as to obtain higher
powers of the small-$k$ behavior, in principle producing any desired
leading behavior. Let us suppose that we generate mass moments
$M^n(\bR)$ at each lattice site $\bR$ defined by
Eq.~(\ref{difference}), with an uncorrelated stochastic process, i.e.,
so that
\begin{equation}
\langle M^n (\bR) M^m (\bR ^\prime) \rangle = 
\langle M^n (\bR) \rangle  \langle M^m (\bR ^\prime) \rangle 
\end{equation}
for each tensor component of the tensor product, for 
$\bR\neq \bR ^\prime$,  (and any $n$, $m$). As in the example given
above starting from Eqs.~(\ref{Sk-calI}) and (\ref{defn-calI}), it
is straightforward to show that $S(\bk)$ may then be written, in a finite 
region around ${\bf k}=0$, as 
\begin{eqnarray}
\label{sk-tensor-GSL}
&&S({\bf k}) = \sum_{n=2}^{\infty} \sum_{m=2}^{\infty}
\frac{(-i)^{m} (i)^{n}}{m!\, n!} \\
&&\times{\bf k}^n \cdot \left[
\langle M^n M^m  \rangle - 
\langle  M^n  \rangle \langle  M^m \rangle \right]
\cdot {\bf k}^m\nonumber
\end{eqnarray}
where all the tensors $M^n$ are evaluated at the same arbitrary
lattice cell, i.e.,
\begin{equation}
[\langle M^n M^m  \rangle]_{\alpha_1...\alpha_n\beta_1...\beta_m}
\equiv 
\langle 
M_{ \alpha_1...\alpha_n} (\bR) M^m  _{ \beta_1...\beta_m} (\bR) \rangle
\end{equation} 
and $k^n \equiv k_{\alpha_1} k_{\alpha_2}...k_{\alpha_n}$ etc..

Thus, for example, we can obtain $\gamma=6$ with an algorithm of this 
type which allows the {\it third} moment $M^3$ of the mass distribution
to vary from cell to cell while keeping the second (quadrupole) moment 
$M^2$ fixed.  This can be done
in a rotationally invariant manner by using a configuration 
of points with a quadrupole moment (relative to its center of mass)
proportional to the identity matrix, e.g., points placed at the 
corners of a regular tetrahedron in $d$ dimensions. Placing
such a configuration at each lattice site, but rotated by a 
random rotation in $SO(d)$, the variance in the second 
moment of the cell is thus zero. However it is easy to verify
that the random rotation engenders a variation in the third moment,
so that one obtains the leading small-$k$ behavior 
$S(\bk) \propto k^6$. The coefficient of 
the $k^6$ term can most easily be made zero by taking 
instead a configuration whose quadrupole moment is again 
diagonal, but whose third mass moment $M^3$ is zero. 
Since the latter is in fact zero for any configuration
of points which is invariant under inversion symmetry,
a possible choice is to add to each tetrahedron a
reflection of itself in its center of mass. Alternatively,
and using fewer points, one can use a randomly rotated
configuration consisting of $d$ couples of points with
equal separation placed orthogonally with common center
of mass. In this case we obtain a leading order small-$k$ 
behavior $S(\bk) \propto k^8$.
Further discussion of this kind of point process generation
can be found in Ref.~\cite{ag-mj-cloudprocesses}. In this
context they arise as a special case of ``cloud processes'',
in which points in a generic initial point process, of which the
correlation properties are assumed known, are replaced by
a ``cloud'' of point particles. 

\subsubsection{Congruent rotationally invariant tilings}

We have seen that superhomogeneous point processes with 
a small-$k$ behavior $S(\bk) \propto k^\gamma$ and $\gamma > 4$ 
can be generated starting from a regular lattice tiling,
by using a stochastic process to determine the positions of
an appropriately constrained set of points in each cell.
While the point process so generated is not statistically
translationally and rotationally invariant, we saw that
a leading behavior proportional to $k$ was obtained
if the stochastic process assigning the points had itself
no preferred direction. Thus at large scales the system 
approximates very well statistical translational and 
rotational invariance.

Starting from a translation and rotation invariant tiling we can
obtain a statistically translation and rotation invariant point
process in an analogous way. It suffices in this case, however, to
assign the points deterministically to each tile, in the same way
relative to each tile (e.g. at the center of mass of the tile). 
We do not need now the additional stochasticity
provided either by taking more points per tile or random positions
inside the tile. The reason is that the moments of the tiles 
given by Eq.~(\ref{mass-moments}) already depend on the tile itself because
of the orientation. Consequently the quantities $\tilde{g}^{n,m}(\bk)$
are non-zero around ${\bf k}=0$ just as in the case when we had one point
per tile. The difference is, as we have discussed at length above,
that we can make certain terms zero so that the leading term appears
at higher order in $k$.

To see how this can be done in a little more detail, let us
consider, to be specific, a pinwheel tiling ($d=2$) or quaquaversal 
tiling ($d=3$) as in  
Sect.~\ref{Congruent rotationally invariant tilings-1} above.
It is convenient to study the SF $S(\bk)$ as given in 
Eq.~(\ref{sk-tensor}), where now the $M^n$ in 
Eq.~(\ref{mass-moments}) are given by the expressions
in Eq.~(\ref{difference}). To define a point distribution in which
the leading term in this expression vanishes, one can proceed
as follows. First one determines the location of the center of mass 
and the quadrupole moment of the elementary tile. From the latter 
one can then find the principal axes, in which it is diagonal. In 
each tile of the tiling one then places on each such axis 
a pair of points, with their center of mass at that of the tile,
at the appropriate distance to produce the component of the second
moment along the corresponding axis. The leading
contribution to $S(\bk)$ should then be determined by the small-$k$
behavior of $\tilde{g}^{3,3} (\bk)$, i.e., by the correlation 
properties of the third moment $M^3$ as in Eq.~(\ref{difference}).
Because of the inversion symmetry in the point distribution in
each tile, the first term in Eq.~(\ref{difference}) vanishes. It
is therefore the correlation properties of the 
third moment of the tiles alone which determines the coefficient 
of the term at order $k^6$. Given the results discussed above
of the numerical studies \cite{Ha06} for the quaquaversal tiling
with a single point at the center of mass, we expect that 
such correlations are short-range, or at most very weakly
long-range. One would expect therefore to obtain an 
$S(k) \propto k^\gamma$ with $\gamma \approx 6$. 

The generalization of this algorithm to higher orders is, 
in principle, straightforward (albeit evidently cumbersome
as the order increases). Further one can seek to determine
the {\it minimal} number of points per tile required to make 
the desired number of terms in Eq.~(\ref{sk-tensor}) vanish.
Indeed the specific algorithm described above uses $2d$ 
points, while it is easy to see that one needs only a smaller 
number to make the leading term in Eq.~(\ref{sk-tensor}) vanish:
in $d=2$, for example, it suffices to have three (rather than
four) points to represent the quadrupole moment of the elementary
triangular tile of the pinwheel tiling. 

\section{Discussion and conclusions}
\label{conclusions}

We have studied a class of point processes generated by placing 
one or more points in each tile of a tiling of $\mathbb{R}^d$.
We have assumed that the tiles have equal volume as this is 
expected to lead to the suppression of fluctuations at large
scales characteristic of superhomogeneous point processes.
We have shown explicitly that one can build 
superhomogeneous point processes with an arbitrarily large
exponent characterizing the small-$k$ behavior of the structure 
factor, and we have presented various examples. 
To our knowledge exact constructions of such point
processes for the case $\gamma > 4$ have not previously been
given in the literature.

We have shown how in these algorithms the exponent $\gamma$ 
depends (i) on the arrangements of the points ascribed to the 
tiles in the algorithm, and (ii) on the correlation properties 
of the shapes and orientations of the tiles. 
For the specific case of regular lattice tilings a
non-trivial contribution to the small-$k$ behavior of $S(\bk)$ 
arises only from the former, with the coefficients in the 
small-$k$ expansion depending explicitly only on the {\it variance} of 
the (tensorial) mass moments of the points assigned to each
cell. By arranging a sufficient number of points in a way which
makes this variance zero for the first $n$ moments, one can 
obtain $\gamma>2n$. In the case of an irregular tiling 
--- we have considered the example of pinwheel and quaquaversal
tilings --- an identical arrangement of the points in each tile
can be sufficient to produce continuous SF and translational and rotational 
invariant superhomogeneous point processes. The exponent
$\gamma$ then encodes information about the correlation properties
of the shapes and orientations of the tiles. If these are
short-range correlated, one obtains $\gamma=4$ placing a single
point at the center of mass of each tile, and $\gamma=2(n+1)$ if
one places a number of points with all mass moments up to the
$n$-th equal to that of the elementary tile. If there is, on
the hand, long-range correlation in the shapes and orientations
of tiles, the exponent obtained will be modified in a way which
depends on the nature of this correlation.

Our results shed light on the meaning of the exponent $\gamma > 0$
characterizing a superhomogeneous point process. Up to the value
$\gamma=4$ previous explicit constructions of discrete processes
(see, e.g., \cite{andrea, gabrielli_etal_04}) have shown 
that the increase of $\gamma$ can be associated with a suppression 
of fluctuations at large scales. Here we have seen that values
$\gamma>4$ correspond indeed to an increased order in the arrangement
of the points, but now {\it at small scales}: it is by changing
how points are arranged within each tile, i.e., below a finite length
scale (but subject always to the global constraints on fluctuations
imposed by the tiling), that we can increase the exponent.  
Thus to ``undo'' the order represented by an exponent $\gamma > 4$ with
respect to a system with $\gamma=4$ requires only the rearrangement 
of the system at small scales, while to ``undo'' that in a system
with $\gamma \leq 4$ require a coherent rearrangement of points 
on arbitrarily large scales (i.e. on scales inverse to the wavenumber 
range in which the exponent is measured).

We have mentioned that our results are relevant in cosmology. Firstly 
the analysis given here makes more rigorous certain heuristic arguments 
used in this context regarding ``causal constraints'' on the generation
of fluctuations from a uniform background \cite{zeldovich-k4,peebles}. An
algorithm like that described here, for the case of a single point 
placed at the center of mass of each tile, has been 
considered \cite{peebles} as a toy model for the generation of 
fluctuations starting from an exactly uniform
mass density, by a physical process which conserves mass and 
momentum {\it locally}. The result $\gamma=4$ is obtained 
by assuming that space is divided into finite cells  
whose positions are uncorrelated. The latter assumption is
in fact not consistent: the division of space into equal volume
cells implies that their positions are necessarily correlated. 
Our more rigorous analysis shows that this exponent $\gamma=4$
does result generically, however, if the shapes and orientations 
of these cells (i.e. tiles) are short-range correlated, i.e.,
have integrable correlation functions. 

Secondly, the generation of very uniform point processes is of
relevance to the generation of initial conditions for numerical
simulations of structure formation in the universe. In this context,
to represent a given set of initial conditions, one must perturb
appropriately  (see \cite{discreteness1_mjbm} for a detailed discussion) 
a point distribution representing as well as possible the uniform
(unperturbed) universe. To understand the effects coming from this
chosen point distribution (which are non-physical) it is 
desirable to have different algorithms which can generate such configurations. 
It is for this
reason that a special case of the algorithm we have studied here has
been built explicitly and studied numerically in this context
\cite{Ha06}.  The analytical results we have given here complement
these studies and give further algorithms for producing even more
uniform point processes which may be useful in this context.  We note
again in this respect that explicit algorithms for producing $\gamma >
4$ have not previously been given. Such distributions in themselves
provide interesting initial conditions (without any perturbation) for
gravitational clustering, which have not previously been studied.

We conclude with some further remarks on our results and some
other directions for further work:

\begin{itemize}

\item In our constructions of point processes we have always
constrained the center of mass of the points in each tile to 
coincide with that of the tile. We have done so because our
goal here has been to generate point processes which are as
uniform as possible. It is a simple exercise to redo our
calculation leading to Eqs.~(\ref{Sk-calI}) and (\ref{defn-calI})
when this constraint is relaxed, i.e., allowing the center of mass 
of the particles (or particle) in each tile to be displaced 
randomly from that of the tile. The result is that the 
leading term in Eqs.~(\ref{Sk-calI}) is now at order $k^2$ 
rather than $k^4$. For short-range correlated tilings
the leading behavior of the SF will then be proportional
to $k^2$. This result will be valid if the displacements of 
the center of mass of the particles within a tile with
respect to the center of mass of the associated tiles 
have a finite variance.  On the other hand, if the
variance of these displacements diverges, the small-$k$ 
behavior of the SF is given by $S(\bk) \propto k^\gamma$ 
where $0<\gamma <2$, the value of the exponent depending 
on the precise behavior displacements PDF for large
arguments. A detailed calculation of these cases for a randomly
perturbed lattice can be found in Ref.~\cite{andrea, book}.

\item While we have shown analytically the existence of
point processes with arbitrarily large exponents $\gamma$,
we have not done so for a case which is statistically 
translation and rotation invariant. In the latter
case our results for the exponent are expressed in terms 
of the small-$k$ behavior of the ${\tilde g}^{n,m} (\bk)$
which encode, as we have explained, information about the 
correlation properties of the shapes and orientations of
the tiles. For the one such example which has been numerically
studied (in Ref.~\cite{Ha06}, the quaquaversal tiling with a single
point at the center of mass) the result indicates an asymptotic
behavior close to (but, as noted in Ref.~\cite{white_WDM}, slightly
different to) that which would arise from a purely short-range 
correlation of the orientations of the tiles. Further numerical 
and analytical study of these points processes would clearly 
be of interest, in particular of the simpler pinwheel tiling.

\item We have made in our derivations here an assumption of 
analyticity at $\bk=0$ of the window function of the tiles,
which corresponds to all moments of their mass distribution 
being finite. This is certainly valid if the tiles are of 
finite extent. It may, however, include other cases which
might be of interest, e.g., in $d>1$ one may envisage that
there is a non-trivial distribution of the shapes of the 
equal volume tiles, in which the extent of a tile is
not limited. One could also consider relaxing the assumption
that the volumes of the tiles are strictly equal, admitting
a distribution of volumes with specified correlation properties.
In analogy with what has been found in certain algorithms 
for $\gamma \leq 4$ \cite{andrea, gabrielli_etal_04}, one 
would expect that such modifications would allow the generation of
point processes with leading non-analytic behavior,
and any value of the exponent $\gamma$. Indeed one would
expect non-analytic exponents to be related either to the divergence
of moments of such a distribution of extent or volume or
to the presence of long-range tile-tile correlations.

\item While all our explicit examples have employed tilings
which are congruent, our results for the small-$k$ behavior 
of the SF $S(\bk)$ all apply only on the much
weaker assumption of equal volume of the tiles. Thus for
example we can apply these results to any tiling generated 
by a deformation of tiles which leaves their volume fixed,
which could encompass a large range of systems of physical 
interest (cells, foams, etc.). We recall in this respect, as 
remarked above, that a point placed randomly in each cell, 
rather than at the center of mass, leads to the restoration
of the order $k^2$ term in the expression given 
in Eqs.~(\ref{Sk-calI}).

\item It is useful to briefly remark on the physical realizability of superhomogeneous
point distributions with arbitrary positive but bounded values of $\gamma$.
Such distributions are disordered to some degree and, although they are unusual,
can be physically constructed. For example, the maximally random jammed (MRJ)
state in three dimensions \cite{To00} is a special disordered sphere packing that
can be regarded to be a prototypical glass because it is perfectly
rigid and yet is maximally disordered. It is a superhomogeneous
point distribution characterized by an exponent
$\gamma=1$ \cite{Do05}, but it is inherently a system
out of equilibrium. While we have examples like the equilibrium one-component plasma
that has $\gamma=2$, can one devise equlibrium superhomogeneous point distributions
in which $\gamma$ is arbitrarily large? The answer is apparently in the affirmative
but it requires more than just pair interactions, namely,
two-, three- and four-body interactions as shown in Ref. \cite{Uc04}.

\end{itemize}


MJ and AG thank ST for hospitality at Princeton University during a
visit in May 2006. MJ thanks B. Jancovici and J. Lebowitz for useful 
discussions. ST gratefully acknowledges the support of the Office of Basic Energy Sciences, DOE, 
under Grant No. DE-FG02-04ER46108. 

\appendix

\section{From Gaussian to spin fields}

\label{appendix}

In this appendix we show explicitly how to generate a regular 
lattice spin configuration with a given two-point correlation function. 
Such a configuration has been used as the starting point in the
RBR algorithm described in Sect.~\ref{Random Binary Rectangular (RBR) tiling}. 

The algorithm we propose is based on a mapping between a set of
correlated Gaussian variables $\{x(\bR)\}$ with zero mean, and the
spin set $\{s(\bR)\}$ (where $\bR$ is, as above, the generic lattice vector).
We do so because to generate a lattice set  
of correlated Gaussian variables with any possible desired
correlation function is very simple.

We denote by
\be
c(\bR)=\left<x(\bR_0)x(\bR_0+\bR)\right>
\label{c}
\ee
and
\[\chi(\bR)=\left<s(\bR_0)s(\bR_0+\bR)\right>\]
the two-point correlation functions of the Gaussian and the spin sets,
respectively. We have used here the statistical lattice translational
invariance. Both $c(\bR)$ and $\chi(\bR)$ must have non-negative
FTs as required by the Khintchine theorem for 
stochastic processes (see e.g. Ref.~\cite{book}).

The starting point is the two variable joint PDF for correlated 
and monovariate Gaussian variables. Denoting by
$x_1$ and $x_2$ two Gaussian variables at two lattice sites 
separated by the vector $\bR$, we have
\bea
p(x_1,x_2;\bR) &&=\frac{1}{2\sqrt{\sigma^4-c^2(\bR)}} 
\nonumber \\
&&\times \exp\left[
-\frac{\sigma^2(x_1^2+x_2^2-2c(\bR)x_1x_2}{2[\sigma^4-c^2(\bR)]}\right]\,,
\eea
where $c(\bR)$ is defined in Eq.~(\ref{c}), and $\sigma^2=c(0)$ is the 
common variance of the Gaussian variables.

The mapping we consider is the simplest possible:
at the site $\bR$ we fix $s(\bR)=1$ if $x(\bR)>0$ and $s(\bR)=-1$ otherwise.
We want now to find the relation between $\chi(\bR)$ and $c(\bR)$.
This can be done simply by noting that we can write
\be
\nonumber
s(\bR)=2\theta[x(\bR)]-1\,,
\ee
 where $\theta(x)$ is the usual Heaviside step function.
Therefore we can write
\bea
\chi(\bR)&&=4\int_{0}^{+\infty}\int_{0}^{+\infty}
\frac{dx_1\,dx_2}{2\sqrt{\sigma^4-c^2(\bR)}} \nonumber \\
&& \times \exp\left[
-\frac{\sigma^2(x_1^2+x_2^2-2c(\bR)x_1x_2}{2[\sigma^4-c^2(\bR)]}\right]
-1\,.
\label{chi}
\eea
Performing the double change of integration variables
\be
\nonumber
\left\{
\begin{array}{l}
y_1=x_1 \\
y_1=x_2-[c(\bR)/\sigma^2]x_1
\end{array}
\right\}
\ee
it is simple to rewrite Eq.~(\ref{chi}) as
\be
\chi(\bR)=2\int_0^\infty\frac{dy_1}{\sqrt{2\pi \sigma^2}}e^{-y_1^2/(2\sigma^2)}\mbox{erf}(Ay_1)\,,
\label{chi2}
\ee
where $A=\frac{c(\bR)}{\sigma\sqrt{2[\sigma^4-c^2(\bR)]}}$ and
\[\mbox{erf}(x)=\frac{2}{\sqrt{\pi}}\int_0^x dt\,e^{-t^2}\] 
is the usual {\em error function}.
We now use the known equality:
\[\int_0^\infty dx\,e^{-px^2}
\mbox{erf}(qx)=\frac{1}{\sqrt{\pi p}}
\mbox{arctg}\left({a\over \sqrt{p}}\right)\]
which implies finally that
\be
\chi(\bR)=\frac{2}{\pi}\mbox{arctg}\left[\frac{c(\bR)}
{\sqrt{\sigma^4-c^2(\bR)}}\right]\,.
\label{chi3}
\ee
Therefore, given a lattice set of monovariate correlated Gaussian variables,
we can map it onto a lattice set of spin variables with $\left<s(\bR)\right>=0$
and $\chi(\bR)$ given by Eq.~(\ref{chi3}).

\subsection{Asymptotics}
From Eq.~(\ref{chi3}) it is simple to verify that
\[\chi(0)=\left<\sigma^2\right>=1\,.\]
Moreover, as for $R=|\bR|\to \infty$ the correlation function $c(\bR)$ 
must vanish, it is simple to verify that for sufficiently large $R$
we have
\[\chi(\bR)\simeq {2\over\pi}{c(\bR)\over \sigma^2}\],
i.e., $\chi(\bR)$ and $c(\bR)$ have the same asymptotic behavior.
In particular if the Gaussian variables are long-/short-range correlated
the spin variables are also long-/short-range correlated with the same
scaling behavior.

We can also give the condition of {\em superhomogeneity} for
the spin lattice set. For the spin system this condition
is simply
\[\sum_{\bR} \chi(\bR)=0\,.\]
which gives the following more complicated relation for 
the correlation function of the Gaussian variables:
\[\sum_{\bR}\frac{2}{\pi}\mbox{arctg}\left[\frac{c(\bR)}
{\sqrt{\sigma^4-c^2(\bR)}}\right] =0\,.\]


\begin{thebibliography}{25}
\expandafter\ifx\csname natexlab\endcsname\relax\def\natexlab#1{#1}\fi
\expandafter\ifx\csname bibnamefont\endcsname\relax
  \def\bibnamefont#1{#1}\fi
\expandafter\ifx\csname bibfnamefont\endcsname\relax
  \def\bibfnamefont#1{#1}\fi
\expandafter\ifx\csname citenamefont\endcsname\relax
  \def\citenamefont#1{#1}\fi
\expandafter\ifx\csname url\endcsname\relax
  \def\url#1{\texttt{#1}}\fi
\expandafter\ifx\csname urlprefix\endcsname\relax\def\urlprefix{URL }\fi
\providecommand{\bibinfo}[2]{#2}
\providecommand{\eprint}[2][]{\url{#2}}

\bibitem[{\citenamefont{Gabrielli et~al.}(2002)\citenamefont{Gabrielli, Joyce,
  and Sylos~Labini}}]{glasslike}
\bibinfo{author}{\bibfnamefont{A.}~\bibnamefont{Gabrielli}},
  \bibinfo{author}{\bibfnamefont{M.}~\bibnamefont{Joyce}}, \bibnamefont{and}
  \bibinfo{author}{\bibfnamefont{F.}~\bibnamefont{Sylos~Labini}},
  \bibinfo{journal}{Phys. Rev.} \textbf{\bibinfo{volume}{D 65}},
  \bibinfo{pages}{083523} (\bibinfo{year}{2002}).

\bibitem[{\citenamefont{Torquato and Stillinger}(2003)}]{To03}
\bibinfo{author}{\bibfnamefont{S.}~\bibnamefont{Torquato}} \bibnamefont{and}
  \bibinfo{author}{\bibfnamefont{F.~H.} \bibnamefont{Stillinger}},
  \bibinfo{journal}{Phys. Rev.} \textbf{\bibinfo{volume}{E68}},
  \bibinfo{pages}{041113} (\bibinfo{year}{2003}), \bibinfo{note}{publisher's
  note: Phys. Rev. E, \textbf{68}, 069901 (2003). Note that the horizontal
axis of Fig. 9 in this paper is mislabeled: $r/(2R)$ should be $r/R$.}

\bibitem[{\citenamefont{Feynman}(1954)}]{Fe54}
\bibinfo{author}{\bibfnamefont{R.~P.} \bibnamefont{Feynman}},
  \bibinfo{journal}{Phys. Rev.} \textbf{\bibinfo{volume}{94}},
  \bibinfo{pages}{262} (\bibinfo{year}{1954}).

\bibitem[{\citenamefont{Feynman and Cohen}(1956)}]{Fe56}
\bibinfo{author}{\bibfnamefont{R.~P.} \bibnamefont{Feynman}} \bibnamefont{and}
  \bibinfo{author}{\bibfnamefont{M.}~\bibnamefont{Cohen}},
  \bibinfo{journal}{Phys. Rev.} \textbf{\bibinfo{volume}{102}},
  \bibinfo{pages}{1189} (\bibinfo{year}{1956}).

\bibitem[{\citenamefont{Reatto and Chester}(1967)}]{Re67}
\bibinfo{author}{\bibfnamefont{L.}~\bibnamefont{Reatto}} \bibnamefont{and}
  \bibinfo{author}{\bibfnamefont{G.}~\bibnamefont{Chester}},
  \bibinfo{journal}{Phys. Rev.} \textbf{\bibinfo{volume}{155}},
  \bibinfo{pages}{88} (\bibinfo{year}{1967}).

\bibitem[{\citenamefont{Donev et~al.}(2005)\citenamefont{Donev, Stillinger, and
  Torquato}}]{Do05}
\bibinfo{author}{\bibfnamefont{A.}~\bibnamefont{Donev}},
  \bibinfo{author}{\bibfnamefont{F.}~\bibnamefont{Stillinger}},
  \bibnamefont{and} \bibinfo{author}{\bibfnamefont{S.}~\bibnamefont{Torquato}},
  \bibinfo{journal}{Phys. Rev. Lett.} \textbf{\bibinfo{volume}{95}},
  \bibinfo{pages}{090604} (\bibinfo{year}{2005}).

\bibitem[{\citenamefont{Baus and Hansen}(1980)}]{OCP-review}
\bibinfo{author}{\bibfnamefont{M.}~\bibnamefont{Baus}} \bibnamefont{and}
  \bibinfo{author}{\bibfnamefont{J.-P.} \bibnamefont{Hansen}},
  \bibinfo{journal}{Physics Reports} \textbf{\bibinfo{volume}{59}},
  \bibinfo{pages}{1} (\bibinfo{year}{1980}).

\bibitem[{\citenamefont{Gabrielli et~al.}(2003)\citenamefont{Gabrielli,
  Jancovici, Joyce, Lebowitz, Pietronero, and Sylos~Labini}}]{lebo_and_all}
\bibinfo{author}{\bibfnamefont{A.}~\bibnamefont{Gabrielli}},
  \bibinfo{author}{\bibfnamefont{B.}~\bibnamefont{Jancovici}},
  \bibinfo{author}{\bibfnamefont{M.}~\bibnamefont{Joyce}},
  \bibinfo{author}{\bibfnamefont{J.~L.} \bibnamefont{Lebowitz}},
  \bibinfo{author}{\bibfnamefont{L.}~\bibnamefont{Pietronero}},
  \bibnamefont{and}
  \bibinfo{author}{\bibfnamefont{F.}~\bibnamefont{Sylos~Labini}},
  \bibinfo{journal}{Phys. Rev.} \textbf{\bibinfo{volume}{D67}},
  \bibinfo{pages}{043506} (\bibinfo{year}{2003}).

\bibitem[{\citenamefont{Joyce et~al.}(2005)\citenamefont{Joyce, 
Levesque, and Marcos}}]{modOCP}
\bibinfo{author}{\bibfnamefont{M.}~\bibnamefont{Joyce}},
  \bibinfo{author}{\bibfnamefont{D.}~\bibnamefont{Levesque}},
  \bibnamefont{and}
  \bibinfo{author}{\bibfnamefont{B.}~\bibnamefont{Marcos}},
  \bibinfo{journal}{Phys. Rev.} \textbf{\bibinfo{volume}{D72}},
  \bibinfo{pages}{103509} (\bibinfo{year}{2005}).

\bibitem[{\citenamefont{Hansen et~al.}(2007)\citenamefont{Hansen, Agertz,
  Joyce, Stadel, Moore, and Potter}}]{Ha06}
\bibinfo{author}{\bibfnamefont{S.}~\bibnamefont{Hansen}},
  \bibinfo{author}{\bibfnamefont{O.}~\bibnamefont{Agertz}},
  \bibinfo{author}{\bibfnamefont{M.}~\bibnamefont{Joyce}},
  \bibinfo{author}{\bibfnamefont{J.}~\bibnamefont{Stadel}},
  \bibinfo{author}{\bibfnamefont{B.}~\bibnamefont{Moore}}, \bibnamefont{and}
  \bibinfo{author}{\bibfnamefont{D.}~\bibnamefont{Potter}},
  \bibinfo{journal}{Astrophys.J.} \textbf{\bibinfo{volume}{656}},
  \bibinfo{pages}{631} (\bibinfo{year}{2007}), \eprint{astro-ph/0606148}.

\bibitem[{\citenamefont{Gabrielli
  et~al.}(2004{\natexlab{a}})\citenamefont{Gabrielli, Sylos~Labini, Joyce, and
  Pietronero}}]{book}
\bibinfo{author}{\bibfnamefont{A.}~\bibnamefont{Gabrielli}},
  \bibinfo{author}{\bibfnamefont{F.}~\bibnamefont{Sylos~Labini}},
  \bibinfo{author}{\bibfnamefont{M.}~\bibnamefont{Joyce}}, \bibnamefont{and}
  \bibinfo{author}{\bibfnamefont{L.}~\bibnamefont{Pietronero}},
  \emph{\bibinfo{title}{Statistical Physics for Cosmic Structures}}
  (\bibinfo{publisher}{Springer}, \bibinfo{year}{2004}{\natexlab{a}}).

\bibitem[{\citenamefont{Zeldovich}(1965)}]{zeldovich-k4}
\bibinfo{author}{\bibfnamefont{Y.~B.} \bibnamefont{Zeldovich}},
  \bibinfo{journal}{Adv. Astron.} \textbf{\bibinfo{volume}{3}},
  \bibinfo{pages}{241} (\bibinfo{year}{1965}).

\bibitem[{\citenamefont{Peebles}(1980)}]{peebles}
\bibinfo{author}{\bibfnamefont{P.~J.~E.} \bibnamefont{Peebles}},
  \emph{\bibinfo{title}{{The Large-Scale Structure of the Universe}}}
  (\bibinfo{publisher}{Princeton University Press}, \bibinfo{year}{1980}).

\bibitem[{\citenamefont{Uche et~al.}(2004)\citenamefont{Uche, Stillinger, and
  Torquato}}]{Uc04}
\bibinfo{author}{\bibfnamefont{O.}~\bibnamefont{Uche}},
  \bibinfo{author}{\bibfnamefont{F.~H.} \bibnamefont{Stillinger}},
  \bibnamefont{and} \bibinfo{author}{\bibfnamefont{S.}~\bibnamefont{Torquato}},
  \bibinfo{journal}{Phys. Rev.} \textbf{\bibinfo{volume}{E70}},
  \bibinfo{pages}{046112} (\bibinfo{year}{2004}).

\bibitem[{\citenamefont{Uche et~al.}(2006)\citenamefont{Uche, Torquato, and
  Stillinger}}]{Uc06}
\bibinfo{author}{\bibfnamefont{O.}~\bibnamefont{Uche}},
  \bibinfo{author}{\bibfnamefont{S.}~\bibnamefont{Torquato}}, \bibnamefont{and}
  \bibinfo{author}{\bibfnamefont{F.~H.} \bibnamefont{Stillinger}},
  \bibinfo{journal}{Phys. Rev.} \textbf{\bibinfo{volume}{E74}},
  \bibinfo{pages}{031104} (\bibinfo{year}{2006}).

\bibitem[{\citenamefont{Gabrielli
  et~al.}(2004{\natexlab{b}})\citenamefont{Gabrielli, Joyce, Marcos, and
  Viot}}]{gabrielli_etal_04}
\bibinfo{author}{\bibfnamefont{A.}~\bibnamefont{Gabrielli}},
  \bibinfo{author}{\bibfnamefont{M.}~\bibnamefont{Joyce}},
  \bibinfo{author}{\bibfnamefont{B.}~\bibnamefont{Marcos}}, \bibnamefont{and}
  \bibinfo{author}{\bibfnamefont{P.}~\bibnamefont{Viot}},
  \bibinfo{journal}{Europhys. Lett.} \textbf{\bibinfo{volume}{66}},
  \bibinfo{pages}{1} (\bibinfo{year}{2004}{\natexlab{b}}),
  \eprint{astro-ph/0303169}.

\bibitem[{\citenamefont{Fratzl et~al.}(1991)\citenamefont{Fratzl, Lebowitz,
  Penrose, and Amar}}]{fratzl_etal1991}
\bibinfo{author}{\bibfnamefont{P.}~\bibnamefont{Fratzl}},
  \bibinfo{author}{\bibfnamefont{J.}~\bibnamefont{Lebowitz}},
  \bibinfo{author}{\bibfnamefont{O.}~\bibnamefont{Penrose}}, \bibnamefont{and}
  \bibinfo{author}{\bibfnamefont{J.}~\bibnamefont{Amar}},
  \bibinfo{journal}{Phys. Rev.} \textbf{\bibinfo{volume}{B44}},
  \bibinfo{pages}{4794} (\bibinfo{year}{1991}).

\bibitem[{\citenamefont{Gabrielli and Torquato}(2004)}]{Ga04}
\bibinfo{author}{\bibfnamefont{A.}~\bibnamefont{Gabrielli}} \bibnamefont{and}
  \bibinfo{author}{\bibfnamefont{S.}~\bibnamefont{Torquato}},
  \bibinfo{journal}{Phys. Rev.} \textbf{\bibinfo{volume}{E70}},
  \bibinfo{pages}{041105} (\bibinfo{year}{2004}).

\bibitem[{\citenamefont{Gabrielli and Joyce}()}]{ag-mj-cloudprocesses}
\bibinfo{author}{\bibfnamefont{A.}~\bibnamefont{Gabrielli}} \bibnamefont{and}
  \bibinfo{author}{\bibfnamefont{M.}~\bibnamefont{Joyce}},
  \emph{\bibinfo{title}{Two point correlation properties of stochastic cloud
  processes}}, \eprint{arXiv:0711.0270}.

\bibitem{To02}
S. Torquato, Random Heterogeneous Materials: Microstructure and Macroscopic Properties
(Springer-Verlag, New York, 2002).

\bibitem[{\citenamefont{Radin}(1995)}]{pinwheel}
\bibinfo{author}{\bibfnamefont{C.}~\bibnamefont{Radin}},
  \bibinfo{journal}{Notices Amer. Math. Soc.} \textbf{\bibinfo{volume}{42}},
  \bibinfo{pages}{26} (\bibinfo{year}{1995}).

\bibitem[{\citenamefont{Conway and Radin}(1998)}]{quaquaversal}
\bibinfo{author}{\bibfnamefont{J.}~\bibnamefont{Conway}} \bibnamefont{and}
  \bibinfo{author}{\bibfnamefont{C.}~\bibnamefont{Radin}},
  \bibinfo{journal}{Inventiones Math.} \textbf{\bibinfo{volume}{132}},
  \bibinfo{pages}{179} (\bibinfo{year}{1998}).

\bibitem[{\citenamefont{Wang and White}()}]{white_WDM}
\bibinfo{author}{\bibfnamefont{J.}~\bibnamefont{Wang}} \bibnamefont{and}
  \bibinfo{author}{\bibfnamefont{S.}~\bibnamefont{White}},
  \emph{\bibinfo{title}{Discreteness effects in simulations of hot/warm dark
  matter}}, \bibinfo{note}{astro-ph/0702575}.

\bibitem[{\citenamefont{Radin}(1999)}]{radin_JSP1999}
\bibinfo{author}{\bibfnamefont{C.}~\bibnamefont{Radin}}, \bibinfo{journal}{J.
  Stat. Phys.} \textbf{\bibinfo{volume}{95}}, \bibinfo{pages}{827}
  (\bibinfo{year}{1999}).

\bibitem[{\citenamefont{Robinson and Wandelt}(1996)}]{wandelt}
\bibinfo{author}{\bibfnamefont{J.}~\bibnamefont{Robinson}} \bibnamefont{and}
  \bibinfo{author}{\bibfnamefont{B.}~\bibnamefont{Wandelt}},
  \bibinfo{journal}{Phys. Rev.} \textbf{\bibinfo{volume}{D53}},
  \bibinfo{pages}{618} (\bibinfo{year}{1996}), \eprint{astro-ph/9507043}.

\bibitem[{\citenamefont{Gabrielli}(2004)}]{andrea}
\bibinfo{author}{\bibfnamefont{A.}~\bibnamefont{Gabrielli}},
  \bibinfo{journal}{Phys. Rev.} \textbf{\bibinfo{volume}{E70}},
  \bibinfo{pages}{066131} (\bibinfo{year}{2004}), \eprint{cond-mat/0409594}.

\bibitem[{\citenamefont{Joyce and Marcos}(2007)}]{discreteness1_mjbm}
\bibinfo{author}{\bibfnamefont{M.}~\bibnamefont{Joyce}} \bibnamefont{and}
  \bibinfo{author}{\bibfnamefont{B.}~\bibnamefont{Marcos}},
  \bibinfo{journal}{Phys. Rev.} \textbf{\bibinfo{volume}{D75}},
  \bibinfo{pages}{063516} (\bibinfo{year}{2007}), \eprint{astro-ph/0410451}.


\bibitem{To00}
S. Torquato, T. M. Truskett, and P. G. Debenedetti, Phys. Rev. Lett. {\bf 84}, 2064 (2000).

\end{thebibliography}

\end{document}